\documentclass[conference]{IEEEtran} 

\usepackage[dvipsnames]{xcolor}

\usepackage[colorlinks=true,linkcolor=blue,breaklinks=True,citecolor=brown,urlcolor=blue]{hyperref}
\usepackage{tabularray}
\usepackage{listings}
\usepackage{dingbat}
\usepackage{ragged2e}
\usepackage{adjustbox}
\usepackage[linesnumbered,algo2e,ruled]{algorithm2e}
\usepackage{multirow}
\usepackage{booktabs,subcaption,dcolumn}
\usepackage{tikz}
\usepackage{enumitem}
\setlist[itemize]{noitemsep, topsep=0pt}
\usepackage{tabularx}
\usepackage[normalem]{ulem}
\usepackage{placeins} 

\usepackage{pifont}
\usepackage{svg}
\usepackage{soul}
\usepackage[font=footnotesize]{caption}
\usepackage[noadjust]{cite}

\usepackage{amsmath}
\usepackage{amsfonts, amssymb}
\usepackage{colortbl}
\usepackage{tablefootnote}
\usepackage[most]{tcolorbox}
\usepackage[sort&compress, numbers]{natbib}

\AtBeginDocument{%
  \providecommand\BibTeX{{%
    \normalfont B\kern-0.5em{\scshape i\kern-0.25em b}\kern-0.8em\TeX}}}

\newcolumntype{?}{!{\vrule width 1.5pt}}

\newtcolorbox{cooltextbox}[1][]{%
    colback=black!5,
    colframe=black!5,
    notitle,
    sharp corners,
    borderline west={0pt}{0pt}{red!80!black},
    enhanced,
    breakable,
    left=0pt,
    right=0pt,
    top=0pt,
    bottom=0pt
    }

\newcommand\revision[1]{%
  \bgroup
  \hskip0pt\color{red!80!black}%
  #1%
  \egroup
}

\newcommand{\quotes}[1]{``#1''}

\begin{document}

\title{Department-Specific Security Awareness Campaigns: A Cross-Organizational Study of HR and Accounting}



\author{
\IEEEauthorblockN{Matthias Pfister\IEEEauthorrefmark{1}, Giovanni Apruzzese\IEEEauthorrefmark{1}\IEEEauthorrefmark{2}, Irdin Pekaric\IEEEauthorrefmark{1}\\}
\IEEEauthorblockA{{ 
\IEEEauthorrefmark{1}\textit{University of Liechtenstein}, Vaduz, Liechtenstein; \IEEEauthorrefmark{2}\textit{Reykjavik University}, Reykjavik, Iceland}
\\
{\small \{matthias.pfister, giovanni.apruzzese, irdin.pekaric\}@uni.li
}}}

\pagestyle{plain}
\maketitle

\begin{abstract}

Many cyberattacks succeed because they exploit flaws at the human level. To address this problem, organizations rely on security awareness programs, which aim to make employees more resilient against social engineering. While some works have, implicitly or explicitly, suggested that such programs should account for contextual relevance, the common praxis in research is to adopt a ``general'' viewpoint. For instance, instead of focusing on department-specific issues, prior user studies sought to provide organization-wide conclusions by treating all participants equally. Such a protocol may lead to overlooking vulnerabilities that affect only specific subsets of an organization, and which can be (or are) exploited by real-world attackers.

In this paper, we tackle such an oversight. First, through a systematic literature review encompassing over 1k papers, we provide factual evidence that prior literature poorly accounted for department-specific needs. Then, building on this (worrying) finding, we carry out a multi-company and mixed-methods study focusing on two pivotal departments of modern organizations: human resources (HR) and accounting. We explore three dimensions: what specific threats are faced by these departments; what topics should be covered in the security-awareness campaigns delivered to these departments; and which delivery methods would maximize the effectiveness of such campaigns for these departments. We begin by interviewing 16 employees of a multinational enterprise, and then use these results as a scaffold to design a structured survey through which we collect the responses of over 90 HR/accounting members of 9 organizations of varying size. We find that HR and accounting departments face distinct threats: HR is targeted through job applications containing malware and executive impersonation, while accounting is exposed to invoice fraud, credential theft, and ransomware. Current training is often viewed as too generic, with employees preferring shorter, scenario-based formats like videos and simulations. These preferences contradict the common industry practice of lengthy, annual sessions. Based on these insights, we propose practical recommendations for designing awareness programs tailored to departmental needs and workflows.

\end{abstract}

\section{Introduction}
\label{sec:introduction}

\noindent
The cybercrime landscape is constantly evolving~\cite{dupont2024broadening_ecrime,sikra2023uk_ecrime,cui2018phishing_ecrime,bruce2024mapping_ecrime}, and automated countermeasures cannot ensure protection against the plethora of threats that may target modern organizations~\cite{fbi2024icr}. While security mechanisms such as firewalls, intrusion detection systems, as well as encrypted communications, still represent a critical component of a secure infrastructure, real-world evidence has shown that most security violations stem from human (mis)behavior~\cite{proofpoint2024phish}. Indeed, despite decades of efforts focused on countering social-engineering attempts~\cite{baki2023sixteen,biggio2018wild,dhamija2006phishing}, the reality is that phishing still works very well~\cite{apwg2024,proofpoint2024phish,barracuda2024report,slashnext2023phishing,talos2024how,alsharnouby2015phishing}. For instance, recent research and technical reports (e.g.,~\cite{proofpoint2024phish,weinz2025impact}) show that attackers now craft their phishing hooks by leveraging large-language models (LLM), or by embedding malicious quick-response (QR) codes in emails---both being means that simultaneously {\small \textit{(i)}}~evade existing ``machine-based'' countermeasures, and {\small \textit{(ii)}}~deceive human users. These threats are further emphasized by the variety of available tools usable to convey phishing attacks~\cite{schroer2025dark,groner2023model}. 

From a technical viewpoint, it is now acknowledged that development of a perfect detection mechanism against cyberattacks is an unfeasible objective~\cite{biggio2018wild}. Given the economics of cybercrime~\cite{akerlof2015phishing,moore2010economics,van2018plug} (which caused \$15+ billions of losses just in 2024~\cite{fbi2024icr}), attackers are highly incentivized to circumvent any automated defense that may be adopted by organizations~\cite{apruzzese2023real}. Therefore, abundant attention has been given to \textit{human-centered} countermeasures---such as phishing education, awareness, and training campaigns~\cite{lain2022phishing, lain2024content, schiller2024employees}. For instance, a 2024 survey across 15 countries and encompassing 1050 security professionals and 7500 end users, revealed that 99\% of the respondents had some sort of security awareness program in their companies~\cite{proofpoint2024phish}. Yet, barely half (53\%) of the respondents asserted that such program was administered to everyone in the respective organization. This finding begs the question ``could it be that the persisting effectiveness of phishing is due to an uneven treatment of phishing education across a company's departments?''

Abundant prior work has shown that such educational campaigns do generally yield more resilient employees~\cite{jampen2020don,jensen2017training}, which translates to an improved cybersecurity culture of the corresponding organization and, hence, potentially leading to an increased protection against social-engineering attacks~\cite{weinz2025impact}. For instance, an employee who quickly reports a phishing email (or talks about receiving a suspicious email~\cite{wash2018provides}) can protect their entire company against the corresponding attack~\cite{sun2024victims}. However, if such a phishing email is sent only to a select number of individuals (e.g., those confined to a specific department of an organization) who did not receive adequate training, then such an oversight may put the entire organization at risk of a data breach~\cite{burton2020wordythief_ecrime}. In other words, what matters is not ``delivering phishing trainings'' (which is allegedly done by almost all organizations~\cite{proofpoint2024phish}), but rather, ``ensuring that such trainings are delivered, and tailored, to all departments---especially those more prone to being targeted by phishing attacks''. Our work is rooted on this principle.

Some research does support such an assertion. For instance, the well-known S\&P'22 and CCS'24 papers by Lain et al. highlight that long-term phishing resilience requires alignment with employees’ actual roles and exposure~\cite{lain2022phishing}, and that phishing training is most effective when the content is aligned with users’ motivations and job functions~\cite{lain2024content}. From a different angle, Burda et al.~\cite{burda2023peculiar} and Meurs et al.~\cite{meurs2022ransomware_ecrime}, show that attackers exploit department-specific communication patterns to craft their phishing emails---indicating that awareness campaigns should be designed to counter such department-specific phishing threats. Even the 2020 extensive literature review in~\cite{jampen2020don} concluded that phishing-training exercises should be adjusted according to the specific employees. And yet, as we will show, the common practice in research is to focus on ``general'' assessments---a habit which does not allow to elucidate department-specific needs and vulnerabilities related to social-engineering threats. This paper seeks to provide evidence of, and then rectify, such a research gap.

\vspace{1mm}

\noindent
\textbf{\textsc{Research Design and Major Findings.}}
There is limited evidence suggesting that prior literature accounted for department-specific issues in cybersecurity awareness campaigns.
Our aim is to propose a set of guidelines and recommendations on how to build targeted cybersecurity campaigns focusing on HR and accounting departments. Let us summarize our work, which encompasses three research activities: 

\begin{itemize}[leftmargin=*]

    \item As a first step, we conduct a systematic literature review by applying keyword-based search and snowballing methodologies. We analyze 1219 papers and identify 29 relevant papers. We find that previous works recognize the importance of tailoring awareness campaigns to the needs of specific audiences. Despite this recognition, organizations still continue to rely on generic awareness campaigns that do \textit{not consider differences} in job roles, existing knowledge levels or cultural contexts. Moreover, existing studies \textit{overlook the importance of real-world organizational context}.
   
    \item We then carry out interviews with 16 employees from HR, accounting, and cybersecurity departments, exploring specific threat perceptions, knowledge gaps, training needs, and delivery preferences. We qualitatively analyze the responses and find that exploitation techniques differ substantially across departmental processes and data types. We also highlight a subtle threat: \textit{malware embedded in fake job applications} delivered via hiring platforms and email attachments---a tactic specifically tailored to HR workflows. Notably, we did not identify any prior literature that explicitly documents or investigates this specific attack vector.

    \item By using our interviews as a scaffold, we design and distribute a structured survey to 93 employees of 9 organizations. Our quantitative analyses reveal statistically significant differences in threat awareness, training format preferences and behavioral barriers between departments. Importantly, we find that participants favored \textit{quarterly training sessions} with video and scenario-driven content. 
    
\end{itemize}
Additional findings demonstrated low awareness of emerging cyber threats. This suggests that training materials are not updated frequently, which is a gap confirmed by participants who acknowledged repetitive content in past sessions. This misalignment with best practices that recommend regular updates could explain the low impact of generic campaigns.
    
Based on our findings, we recommend shifting from a \quotes{one-size-fits-all} to a \quotes{fit-for-context} model of cybersecurity awareness. Our findings are particularly relevant for practitioners in security operations, risk management, and law enforcement, as well as policymakers aiming to foster secure digital behavior in critical sectors. Thus, this work bridges the gap between academic research and practice by offering evidence-based recommendations to enhance the efficacy of cybersecurity awareness campaigns.









\vspace{1mm}

\noindent
\textbf{\textsc{Contributions.}} We contribute to cybercrime research by:
\begin{itemize}[leftmargin=*]
    \item showing that prior work on cybersecurity awareness overlooked department-specific issues of modern organizations;
    \item carrying out a mixed-methods and multi-organizational user study focusing on various aspects of cybersecurity awareness programs in HR and accounting departments:
    \item providing recommendations based on our original findings.
\end{itemize}
For transparency, we provide low-level details of our research methods (e.g., questionnaire and interview guides)~in~the~Appendix. We also share additional resources in our repository: \url{https://github.com/irdin-pekaric/eCrime2025}.
\section{Related Work and Motivation}
\label{sec:related}

\noindent
We summarize cybersecurity awareness programs, and briefly position HR and accounting in the~cybercrime landscape~(§\ref{ssec:background}). Then, we present our systematic literature review~(§\ref{ssec:slr}), and define the research gap~(§\ref{ssec:gap}).


\subsection{Background}
\label{ssec:background}

\noindent 
Organizations invest a massive amount of resources in cybersecurity measures to defend against a diverse range of threats~\cite{dupont2019cyber_ecrime,sikra2023uk_ecrime}. 
However, it has been shown that currently-deployed ``machine-based'' countermeasures can be evaded easily (e.g.,~\cite{acharya2021phishprint_ecrime,timko2023commercial_ecrime,weinz2025impact}). Therefore, cybersecurity awareness programs now represent a pivotal asset in the overall security infrastructure of modern organizations~\cite{vasudevan_decade_2023,schiller2024employees, sauerwein2019analysis}.


\subsubsection{\textbf{Cybersecurity Awareness Campaigns}}
\label{sssec:cac}

These activities\footnote{We use the term ``awareness'', but other terms exist, such as ``training'' and ``education.'' From a semantical standpoint, the ``awareness'' is necessary to both ``training'' and ``education'' (i.e., you cannot ``educate'' or ``train'' an employee if they are not ``aware'' of a given threat~\cite{wilson_information_1998}) which is why we will use the term ``awareness'' in the remainder of this paper.}
seek to inform and educate employees on various cybersecurity threats, with the purpose of enabling employees to recognize and respond to malicious actions, policy violations and other security-noteworthy events~\cite{vasudevan_decade_2023}. 
Traditional awareness programs rely on a variety of approaches, including newsletters, static e-learning modules, or even attack simulations~\cite{weinz2025impact,hillman2023evaluating,darem2021anti}. It has been found that employees undergoing such awareness programs do indeed improve their security attitude~\cite{lain2022phishing,caldwell2016making,hwang2021security}. However, various prior work have also criticized similar approaches, since they can lead to a false-sense of security (e.g., employees may show improvements in ``tests'', but may still fail ``in practice''~\cite{schiller2024employees}), and may lead to a reduced productivity (or increased stress), since employees must tend to such educational activities on top of the tasks pertaining their actual role in an organization~\cite{chua2019identifying_ecrime}. Put simply, despite their potential effectiveness, security awareness programs present much room for improvement.





\subsubsection{\textbf{Threats to HR and Accounting}} 
\label{sssec:threats}
These departments have a strategic role in modern organizations due to their access to personal and financial data. As such, they often serve as gateways to sensitive assets, information, and communication flows---making these departments rather attractive targets for attackers. Social engineering techniques -- which can trick employees into revealing credentials, downloading malicious attachments or authorizing fraudulent transactions -- often target HR workflows that include recruitment or employee onboarding, which exposes staff to external communications~\cite{williams_why_2024}. For instance, attackers may leverage fictitious resumes to impersonate applicants~\cite{oxford_college_of_management_crucial_2024}. 
In parallel, the workflows of accounting departments (such as vendor payments and payroll processing) are particularly prone to exploitation due to their urgency and perceived authority. Common well-known techniques include Business Email Compromise (BEC), invoice fraud and CEO impersonation scams~\cite{williams_top_2025, sentinelone_cyber_2025}, which often bypass technical controls by focusing on process-specific vulnerabilities as well as trust relationships within teams. 




\subsection{Systematic Literature Review}
\label{ssec:slr}

\noindent
As a starting point of our research, we asked ourselves a preliminary research question (RQ0): 
\textit{\textcolor{violet!90!black}{``to what extent has prior research on security awareness campaigns accounted for department-specific issues in the examined organizations?''}} We tackle RQ0 via a systematic literature review (SLR).

\subsubsection{\textbf{Methodology}}
\label{sssec:slr_method}
Our SLR follows established practices: we first collect relevant papers, and then analyse them~\cite{rethlefsen2021prisma}, in line with prior systematic reviews in security domains~\cite{pekaric2023systematic}. In what follows, we describe the systematic protocol we adopted to carry out each of these phases (carried out in Q1 2025).

\begin{itemize}[leftmargin=*]
    \item \textit{Collection phase.} Following the guidelines by Kitchenham et al.~\cite{kitchenham2015evidence} and Webster and Watson~\cite{webster2002analyzing}, we combine structured keyword-based searches with backward and forward snowballing. First, we designed two sets of query strings,\footnote{We centered the queries around the \quotes{cybersecurity} and \quotes{awareness} keywords. These were then paired with complementary terms such as \quotes{program}, \quotes{campaign} and \quotes{department} (including their synonyms) to refine the scope and align the search more closely with our research objectives. Due to some digital libraries returning too many irrelevant results, two separate search strings were crafted. We discuss the limitations of our search queries in §\ref{ssec:limitations}} a ``short'' (in Listing~\ref{lst:short}) and a ``long'' (in Listing~\ref{lst:long}) string, which we used to search for papers (published after 2015) indexed in five popular literature databases: \textit{ACM Digital Libraries}, \textit{IEEE Xplore}, \textit{Elsevier ScienceDirect}, \textit{SpringerLink}, \textit{Association for Information Systems} (AIS); and we also included the proceedings of the \textit{USENIX Security Symposium} and \textit{NDSS Symposium} (see Appendix \ref{sapp:slr}). Overall, this search yielded 1219 unique papers. Next, we manually inspected the titles and abstracts of these papers, filtering results outside our scope; for instance, we excluded papers proposing technical countermeasures (e.g.,~\cite{humaidi2022procedural,hart2020riskio}) or theoretical security models (e.g.,~\cite{safa2016information,safa2016informationa}), or literature reviews/systematizations (e.g.,~\cite{jampen_dont_2020,sharif_review_2020, pahlavanpour_systematic_2024} or that were not related to cybersecurity (e.g.,~\cite{barreda2015generating,dabbous2020bridging}), or that were not in English (e.g.,~\cite{mendoza2020concientizacion,infante2022factores}). Such a filtering process led to 21 papers being within our scope.\footnote{For the sake of clarity, our focus is on ``papers that explicitly carried out user studies on security awareness''; by ``user studies'', we mean either interviews, or user surveys, or simulations (similarly to~\cite{pekaric2025we}).} Then, we applied forward and backward snowballing, looking for all works that cited, or were cited by, such 21 papers (and that were within our scope); we found 6 more papers in this way. Finally, we decided to also include 2 white papers that we found by issuing the ``short'' string on Google, leading to 29 papers in total. All these operations have been carried out by two authors who interacted to resolve uncertainties.

    
    \item \textit{Analysis phase.} The 29 papers have been manually analysed. The primary goal of such an analysis was to answer our preliminary RQ, which was done by assessing if the study incorporated department-specific or role-based perspectives. However, we also assessed additional aspects, such as: considered organizations, sample size, methodology, and objectives. To this purpose, we developed a codebook (which we report in the repository) through which two authors analysed each paper. We also measured the inter-code reliability: Cohen $\kappa$=0.94, indicating strong agreeability~\cite{warrens2015five}.    
\end{itemize}
We report in Table~\ref{tab:slr} the results of coding-based analysis for each paper found in the collection phase.



\begin{lstlisting}[caption={Short Query},label={lst:short}]
("security" OR "cybersecurity") AND ("awareness") AND ("program" OR "campaign") AND ("organization" OR "company" OR "department" OR "corporation" OR "enterprise")
\end{lstlisting}

\begin{lstlisting}[caption={Long Query. Some keywords were applied only on the Abstract},label={lst:long}]
("security" OR "cybersecurity") AND ("awareness") AND (Abstract:("program") OR ("campaign")) AND (Abstract:("organization") OR Abstract:("company") OR Abstract:("department") OR Abstract:("corporation") OR Abstract:("enterprise"))
\end{lstlisting}

\subsubsection{\textbf{Results}}
\label{sssec:slr_findings}

Let us highlight the major findings stemming from Table~\ref{tab:slr}, and pertaining to our preliminary RQ. 
\begin{itemize}[leftmargin=*]
    \item \textit{General vs Specific.} The wide majority (82\%) of reviewed studies conceptualize cybersecurity awareness campaigns as \quotes{universal interventions} that are intended for the general employee population (e.g.,~\cite{alshaikh_applying_2021, sterk_it_2021}). These campaigns typically cover topics such as phishing, password hygiene and email security. However, these studies do not consider domain-specific workflows, risk exposure or operational contexts. We note that only 5 out of 29 ($\approx17\%$) papers made an explicit reference to role- or department-specific tailoring (i.e.,~\cite{abu-amara_cyber_2021, alkhazi_assessment_2022, berens_taking_2024, gamisch2023study, bauer_prevention_2017}). However, only a single paper~\cite{rampold_custom_2024} investigated the process of crafting targeted campaigns. However, this was done in a healthcare domain and the focus was not on departments, but employees' roles.
    
    \item \textit{Real-world vs Proxies.} Most empirical studies ($\approx76\%$) were conducted using student populations, simulated environments or convenience samples from single organizations (e.g.,~\cite{gamisch2023study,he_effects_2017,reeves_get_2021}). While these designs offer controlled insights, they do not capture the complexity of real-world organizational dynamics. 
    
    \item \textit{Single vs Multi-organization.}
    Also, most studies ($\approx78\%$) only pertain to a single organization, making it hard to determine the extent to which any finding may hold across different organizations. For instance, Nicholson et al. \cite{nicholson_introducing_2018} conducted their study within a single organization, while Alshaikh et al. \cite{alshaikh_exploratory_2018} and Khando et al. \cite{khando_enhancing_2021} conducted research across multiple organizations.
\end{itemize}
Based on these quantitative findings, we can answer RQ0.

\begin{cooltextbox}
    \textsc{\textbf{Answer to RQ0:}}
    Prior work on cybersecurity awareness programs has poorly accounted for department-specific issues. Only one paper (out of 29) considered carrying out targeted campaigns---but this was done on the basis of an employee's role, and not of their department.
\end{cooltextbox}

\subsubsection{\textbf{Observations}}
\label{sssec:slr_observations}

We enrich the findings of our SLR by making additional low-level observations. First, although some studies investigate and compare delivery formats (such as video-based~\cite{reeves_get_2021,abawajy_user_2014,berens_taking_2024,he_effects_2017,sterk_it_2021} vs. text-based content~\cite{ alkhazi_assessment_2022, abu-amara_cyber_2021, carella_impact_2017}, or microlearning~\cite{kavrestad_evaluation_2022} vs. game-based~\cite{alyami_critical_2022, alkhazi_assessment_2022, abu-amara_cyber_2021}) no paper examined whether training effectiveness varied across departments or job types. This is problematic given the increasing diversity of communication styles, workload patterns and cognitive requirements across roles. Second, even though six papers~\cite{rampold_custom_2024,alshaikh_applying_2021, alkhazi_assessment_2022, nicholson_introducing_2018, he_effects_2017, ki-aries_persona-centred_2017} do make some department-specific consideration, terms such as ``human resources'' or ``accounting'' (or even ``finance'') are never mentioned\footnote{These terms appear in~\cite{nicholson_introducing_2018}, but only to specify that ``support participants included receptionists and staff in accounting and HR departments.''} in the text---indicating a lack of research that specifically focuses on these two crucial departments in modern organizations. Third, and as a disclaimer: \textit{our SLR is not meant to invalidate prior work}. We simply point out an (arguably) crucial aspect of security-awareness programs that has been overlooked.

\begin{table}[ht]
\caption{Our Classification Table}
\label{tab:slr}
\footnotesize
\centering
\begin{tabularx}{\linewidth}{
  @{}|@{\hspace{0pt}}>{\centering\arraybackslash}p{1.5cm}@{\hspace{0pt}}|
  @{\hspace{0pt}}>{\centering\arraybackslash}p{0.6cm}@{\hspace{0pt}}|
  @{\hspace{0pt}}>{\centering\arraybackslash}p{1.3cm}@{\hspace{0pt}}|
  @{\hspace{0pt}}>{\centering\arraybackslash}p{1.5cm}@{\hspace{0pt}}|
  @{\hspace{0pt}}>{\centering\arraybackslash}p{3cm}@{\hspace{0pt}}|
  @{\hspace{0pt}}>{\centering\arraybackslash}X@{\hspace{0pt}}|@{}
}
\hline
\textbf{Paper} & \textbf{Year} & \textbf{Source} & \textbf{Type of User Study} & \textbf{Focus} & \textbf{Sample Size} \\
\hline
Rampold et al. \cite{rampold_custom_2024} & 2024 & HICSS & Survey + Interviews & Secure behavior of personnel in healthcare & 8 + 12 \\ \hline
Berens et al. \cite{berens_taking_2024} & 2024 & COSE & Field Study  & Effectiveness of cybersecurity training for older adults & 35 \\ \hline
Schöni et al. \cite{schoni_you_2024} & 2024 & EuroUSEC & Field Study & Evaluation of phishing training & 122  \\ \hline
Chowdhury et al. \cite{chowdhury_personalized_2023} & 2023 & IJIS & Survey + Interviews & Cybersecurity Training for non-IT personnel & 113 + 10 \\ \hline
Gamisch and Pöhn \cite{gamisch2023study} & 2023 & ARES & Field Study & Comparison of multiple awareness campaigns & 261\\ \hline
Hull et al. \cite{hull_tell_2023} & 2023 & COSE & Online Study & Influence of storytelling on security behavior & 339 \\ \hline
Keshvadi et al. \cite{keshvadi_enhancing_2023}& 2023 & IHTC &  Survey + Training & Cybersecurity Training for non-IT personnel & 208 + 12 \\ \hline
Reeves et al. \cite{reeves_generic_2023} & 2023 & COSE & Interviews + Field Study & Employee perceptions of organizational security & 32\\
\hline
Tally et al. \cite{tally_what_2023} & 2023 & TOCHI & Survey + Interviews & Cybersecurity perceptions of mid-career personnel & 33 + 101 \\ \hline
F. Adilia \cite{adilia2023instagram} & 2023 & ISCTE-IUL & Survey + Interviews & cybersecurity awareness campaign via Instagram & N/A \\ \hline
Alkhazi et al. \cite{alkhazi_assessment_2022}& 2022 & IEEE Access & Survey & Impact of awareness programs on behavior & 420 \\ \hline
Alyami et al. \cite{alyami_critical_2022} & 2022 & Cyber-RCI & Survey & Success factors for SETA programs & 153  \\ \hline
Chowdhury et al. \cite{chowdhury_modeling_2022}& 2022 & COSE & Surveys & Cybersecurity training for non-IT professionals & 113 \\ \hline
Kävrestad et al. \cite{kavrestad_evaluation_2022}& 2022 & Future Internet & Experiment & Contextual vs. game- based phishing methods & 92  \\ \hline
Li et al. \cite{li_effects_2022}& 2022 & CHBR & Survey & Compliance with security policies & 389 \\ \hline
H. Awaludin \cite{awaludin2022instagram} & 2022 & Malmö University & Survey + Interviews & Awareness via Instagram among students & 40 \\ \hline
Abu-Amara et al. \cite{abu-amara_cyber_2021}& 2021 & INDIA
Com & Pre/Field Study & Gamified cybersecurity awareness program & 10 \\ \hline
Alshaikh et al. \cite{alshaikh_applying_2021}& 2021 & COSE & Field Study & Evaluation of existing awareness approaches  & 24  \\ \hline
Reeves et al. \cite{reeves_get_2021} & 2021 & COSE & Interviews & Perceptions of security training content & 26  \\ \hline
Sterk und Heinemann \cite{sterk_it_2021} & 2021 & EICC & Field Study & Evaluation of board game for teaching security & 32 \\ \hline
Rege et al. \cite{rege_social_2020} & 2020 & ISEC & Field Study & Social engineering training for students  & N/A \\ \hline
Schütz and Fertig \cite{schutz_analyze_2020} & 2020 & HICSS & Survey &  Behavior model for security compliance & 204 \\ \hline
Alshaikh et al. \cite{alshaikh_exploratory_2018} & 2018 & HICSS & Interviews & Exploration of security awareness practices & 18  \\ \hline
Nicholson et al. \cite{nicholson_introducing_2018}& 2018 & USENIX & Experiment & Security decision-making using Cybersurvival & 111 \\ \hline
Anjaria and Mishra \cite{anjaria_neuroevolution_2017} & 2017 & IC SoftComp & Survey & Neuroevolution adaptive training & 145 \\ \hline
Carella et al. \cite{carella_impact_2017} & 2017 & IEEE Big Data & Field Study & Effects of awareness training on phishing & 753  \\ \hline
He et al. \cite{he_effects_2017} & 2017 & AMCIS & Experiment & Comparison of interactive training methods & 228 \\ \hline
Ki-Aries and Faily \cite{ki-aries_persona-centred_2017} & 2017 & COSE & Interviews & Development of persona awareness training & 9 \\ \hline
Bauer et al. \cite{bauer_prevention_2017} & 2017 & COSE & Interviews & Design of awareness programs for banks & 33 \\

\hline
\end{tabularx}
\vspace{-3mm}
\end{table}

\subsection{Research Gap and Problem Statement}
\label{ssec:gap}

\noindent
Our SLR highlighted that, despite the existence of dozens of papers involving, explicitly or implicitly, cybersecurity-awareness programs, limited attention has been given to department-specific issues---especially those affecting the HR and accounting departments (§\ref{sssec:slr_findings}). 

To rectify such an oversight, we make the first step and carry out a novel user study focused specifically on these two departments of various organizations. Our study is designed to answer the following three RQ:

\begin{itemize}[leftmargin=1cm]
    \item[RQ1:] \textcolor{violet!90!black}{\textit{What are the specific cybersecurity threats faced by HR and accounting departments?}} (§\ref{ssec:results_1})

    \item[RQ2:] \textcolor{violet!90!black}{\textit{What topics should cybersecurity awareness campaigns focus on to maximize their effectiveness for HR and accounting departments?}} (§\ref{ssec:results_2})

    \item[RQ3:] \textcolor{violet!90!black}{\textit{Which delivery methods are effective for engaging with HR and accounting employees in cybersecurity awareness campaigns?}} (§\ref{ssec:results_3})
\end{itemize}
Our RQs are inspired by the findings discussed in §\ref{sssec:slr_observations}. We note that such RQ are particularly valuable ``today'', given that they provide novel insights pertaining to HR and accounting departments in the current cybercrime landscape.

\section{Research Method}
\label{sec:method}

\noindent
To answer our research questions, we conducted semi-structured interviews (§\ref{ssec:method1}) and user surveys (§\ref{ssec:method2}). Interviews were conducted only with a case company, while surveys targeted nine different companies. Both of these were developed in agreement with the case company. Additionally, we shared our survey on a well-known professional online social network, LinkedIn, to serve as a reference point for validating the results obtained from company responses (§\ref{ssec:method3}). 

\subsection{Semi-structured Interviews}
\label{ssec:method1}
\noindent 
In the first phase of the study, we conducted semi-structured interviews with employees and managers from HR, accounting, and cybersecurity departments to explore threats, training content needs, and delivery preferences specific to each department. With this method~\cite{kallio_systematic_2016}, we investigated gaps and insights identified in our systematic literature review (§\ref{ssec:slr}), while also addressing emerging insights obtained during interviews. 

\subsubsection*{Design}
We defined two participant groups: (1) cybersecurity specialists and (2) HR and accounting department employees and managers. We included cybersecurity specialists to ensure technical precision in threat identification, as non-technical personnel may lack the expertise to fully recognize or articulate complex attack scenarios~\cite{wilson_information_1998}. We developed interview guides for each target group based on Kallio et al.'s~\cite{kallio_systematic_2016} five-phase approach, incorporating themes derived from a systematic literature review. Themes included tailoring awareness campaigns, training delivery methods, and behavior change. These were structured using principles from Protection Motivation Theory \cite{boer1996protection}. Following the guidance of Turner~\cite{turner_research_2017} and Venkatesh et al.~\cite{venkatesh_bridging_2013}, the interview guides included a warm-up phase to collect demographic and role specific information, a main section to discuss department-specific threats, training topics, delivery methods, and content relevance, targeting the needed aspects to answer the defined RQ, and a cool-down phase to facilitate open-ended reflection. Pre-tests were conducted with two representatives from each group to ensure clarity, flow, and approximate duration, following the guidelines by Helfferich~\cite{helfferich_leitfaden-_2019}. A 30-minute target was established for the duration of the interview. The final interview guides contained 15 questions for cybersecurity specialists and 14 for non-technical participants (see Tables \ref{tab:interview_guide_departments} and \ref{tab:interview_guide_cyber_specialists}).

\subsubsection*{Distribution}
A total of 16 interviews were conducted (see Table \ref{tab:interview_data}): Four were with security professionals, and twelve were with departmental employees (six each from HR and accounting). Participants were selected by purpose-sampling to ensure domain relevance, reliability, and availability, in accordance with Gläser and Laudel's~\cite{glaser_experteninterviews_2009} criteria. All interviews were conducted remotely via Microsoft Teams during calendar weeks 7 and 8 of 2025. They were recorded with participant consent, transcribed, and anonymized for analysis.

\begin{table}[htbp]
\caption{Interview Guide for HR and accounting departments}
\label{tab:interview_guide_departments}
\footnotesize
\centering
\begin{tabularx}{\columnwidth}{|c|c|c|X|}
\hline
\textbf{ID} & \textbf{RQ} & \textbf{Category} & \textbf{Question} \\
\hline
1 & - & Warm-Up & What is your current role? \\
\hline
2 & - & Warm-Up & How much experience do you have in your role at this company? \\
\hline
3 &  - & Warm-Up & How old are you? \\
\hline
4 & 1 & Warm-Up & How well would you consider your knowledge of cyber security? \\
\hline
5 & 1 & Topics & Have you participated in any cyber security awareness programs before? \\
\hline
6 & 2 & Topics & Which topics did you find most useful or relevant? \\
\hline
7 & 2 & Topics & Are there any topics you think are underrepresented in current cyber security awareness programs? \\
\hline
8 & 3 & Delivery Methods & Which types of cyber security awareness materials (e.g., videos, workshops, email campaigns, games) do you find most engaging? \\
\hline
9 & 3 & Delivery Methods & What changes would you suggest to make cyber security awareness campaigns more relevant and engaging for your department? \\
\hline
10 & 3 & Delivery Methods & Do you think short and frequent training sessions (micro-learning) would be effective for your department? \\
\hline
11 & 2 & Content & Are there particular examples or scenarios that would make cyber security training more relatable to your job responsibilities? \\
\hline
12 & 2 & Content & Do you think the content of awareness campaigns should be tailored to specific roles within your department? \\
\hline
13 & - & Cool-Down & Is there anything else you would like to add that was not covered in this interview that you think is important? \\
\hline
14 & - & Cool-Down & Do you have a question for me? \\
\hline
\end{tabularx}

\end{table}

\begin{table}[ht]
\caption{Participant and Interview Characteristics}
\label{tab:interview_data}
\footnotesize
\centering
\begin{tabularx}{\linewidth}{|X|c|X|c|X|c|}
\hline
\textbf{ID} & \textbf{Department} & \textbf{Age} & \textbf{Experience} & \textbf{Duration} \\
\hline
CS1 & Cyber Security & 35-44 & 13 & 00:28:15 \\
CS2 & Cyber Security & 25-34 & 2 & 00:28:55 \\
CS3 & Cyber Security & 35-44 & 9 & 00:26:43 \\
CS4 & Cyber Security & 18-24 & 2 & 00:29:07 \\
A1 & Accounting & 45-54 & 20+ & 00:22:25 \\
A2 & Accounting & 25-34 & 3 & 00:24:09 \\
A3 & Accounting & 45-54 & 10+ & 00:24:47 \\
A4 & Accounting & 45-54 & 12 & 00:29:30 \\
A5 & Accounting & 25-34 & 4 & 00:25:36 \\
A6 & Accounting & 45-54 & 16 & 00:28:48 \\
HR1 & Human Resources & 25-34 & 2 & 00:24:21 \\
HR2 & Human Resources & 25-34 & 1 & 00:42:04 \\
HR3 & Human Resources & 25-34 & 2 & 00:42:04 \\
HR4 & Human Resources & 25-34 & 8 & 00:30:00 \\
HR5 & Human Resources & 25-34 & 4 & 00:23:56 \\
HR6 & Human Resources & 45-54 & 20+ & 00:25:12 \\
\hline
\end{tabularx}

\end{table}

\subsubsection*{Data Analysis}
We analyzed the transcripts using a deductive content analysis approach grounded in the methodology of Mayring~\cite{mayring_qualitative_2010}. Eight thematic categories were defined in advance based on findings from our systematic literature review (§\ref{ssec:slr}): (1) perception of cybersecurity awareness campaigns, (2) department-specific threats, (3) knowledge gaps and training needs, (4) delivery method preferences, (5) behavior and organizational culture, (6) barriers and challenges, (7) customization and personalization preferences, and (8) measurement and feedback mechanisms. Responses were segmented into relevant themes to enable a structured comparison across departments. This allowed for the identification of shared patterns and department-specific needs. Insights gained from this analysis informed the development of a survey instrument for the next phase of the study.

\subsection{Company Surveys}
\label{ssec:method2}

\noindent The second phase of the study aimed to validate and extend the findings from the qualitative interviews (§\ref{ssec:method1}) by conducting a quantitative survey. This approach follows the triangulation model by Turner et al.~\cite{turner_research_2017} and the logic-based mixed-method framework of Venkatesh et al.~\cite{venkatesh_bridging_2013}. The survey targeted employees and managers of HR and accounting departments. 

\subsubsection*{Design}
Unlike the interview phase, which included cybersecurity specialists, this phase exclusively focused on non-technical staff to assess their perceived risks, training preferences, and campaign relevance. This targeted sampling increased the study's internal validity and enabled cross-sectional comparisons between departments. The development of the survey was based on the methods proposed by Döring~\cite{doring_forschungsmethoden_2023} and Taherdoost~\cite{taherdoost_data_2021}. The Questions were grouped into five thematic blocks: (1) demographics and context, (2) current training, (3) knowledge and threat awareness, (4) delivery method preferences, and (5) organizational culture and management support. Each block was constructed to address specific aspects of cybersecurity awareness, directly derived from the findings of previous phases. The \textit{Demographics and Context} section includes background information about the participants, such as their department affiliation, role, and years of experience. \textit{Current Training} section addresses the participants’ prior experience with cybersecurity training, including the training’s format, frequency, and perceived effectiveness. This information was used to determine baseline training saturation and individual experiences with existing campaigns. The majority of the survey was allocated to the \textit{Knowledge and Threat Awareness} section that aims to evaluate employees’ knowledge of specific cybersecurity risks relevant to their role, such as phishing,
invoice fraud, and social engineering. It also includes items measuring self-assessed readiness and confidence in threat detection. For this part, we used relevant topics from the HAIS-Q questionnaire as the construct has already been validated by previous research~\cite{PARSONS201740}. In addition, both \textit{Current training Knowledge} and \textit{Threat Awareness} sections were structured around the Protection Motivation Theory. Section \textit{Delivery Method Preferences} aims to evaluate the design of customized campaigns, including preferences for training formats, delivery channels, and scheduling. \textit{Organizational Culture and Management Support}  section examined how participants perceive cybersecurity as a shared organizational responsibility. The focus was on the commitment of the management, the support of the department, and the perceived severity of the cybersecurity communication within the teams. Questions were closed-ended (yes/no, Likert scale, and ranking) to ensure reliability and enable quantitative comparisons. The survey was implemented using the GDPR-compliant Swiss platform \quotes{Findmind} (\url{https://en.findmind.ch/}). A pretest with six participants confirmed the clarity of the questions and usability of the tool, and adjustments were made based on the feedback received. The final questionnaire had 36 closed-ended items and was available in English and German to improve accessibility (see Table \ref{tab:survey_questions_answers}). 

\begin{table}[ht]
\caption{Company Characteristics and Response Participation}
\label{tab:company_responses}
\footnotesize
\centering
\begin{tabularx}{\linewidth}{|X|r|X|c|}
\hline
\textbf{Company} & \textbf{Employees} & \textbf{Sector} & \textbf{Responses} \\
\hline
Case Company & 34,000 & Manufacturing & 30 \\
1 & 1,000 & Banking & 19 \\
2 & 1,300 & Healthcare & 14 \\
3 & 500 & IT & 10 \\
4 & 2,000 & E-commerce & 9 \\
5 & 2,000 & Industry & 2 \\
6 & 20 & Industry & 1 \\
7 & 88,000 & Industry & 7 \\
8 & 7,000 & IT & 1 \\
\hline
\end{tabularx}

\end{table}

\subsubsection*{Distribution}
The recruitment process was multi-stage. In this case, the survey was distributed internally at the case company. Additionally, eight external organizations from sectors including banking, healthcare, and IT participated. While the primary focus was on HR and accounting departments, we also included participants from other departments within the surveyed organizations. This \quotes{Other} group serves only to better contextualize department-specific differences between HR, accounting, and other departments: including these respondents allowed us to distinguish whether observed patterns were truly specific to HR and accounting. To protect respondent anonymity, distribution was decentralized via designated company links. Data collection took place between calendar weeks 17 and 22 of 2025. A total of 93 participants completed the survey: 30 came from the case company and 63 came from external organizations.  A break down of the participants of this user study is in Table~\ref{tab:company_responses}. In terms of their role, 43 participants belong to HR, 44 to accounting, and 6 to ``other'' departments. More information about the demographics of our survey is provided in the Appendix~\ref{sapp:demographics}.

\begin{figure*}[!t]
    \centering

    \begin{subfigure}[t]{0.32\textwidth}
        \centering
        \includegraphics[width=\textwidth]{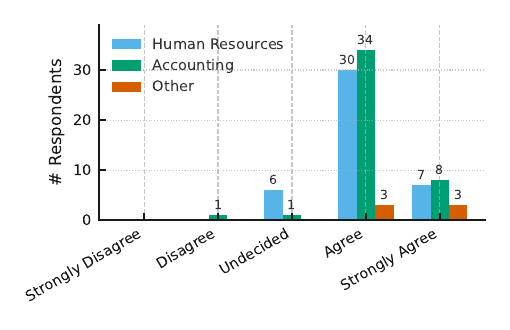}
        \caption{Recognition of department-specific phishing threats}
        \label{fig:phishing}
    \end{subfigure}
    \hfill
    \begin{subfigure}[t]{0.32\textwidth}
        \centering
        \includegraphics[width=\textwidth]{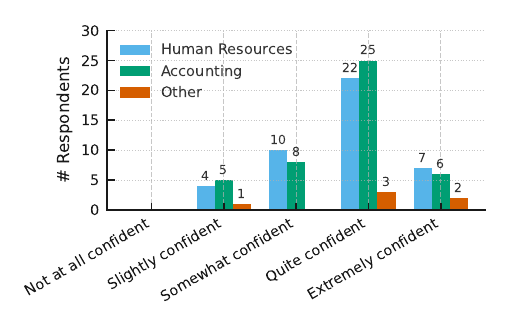}
        \caption{Confidence in identifying cybersecurity threats}
        \label{fig:confidence}
    \end{subfigure}
    \hfill
    \begin{subfigure}[t]{0.32\textwidth}
        \centering
        \includegraphics[width=\textwidth]{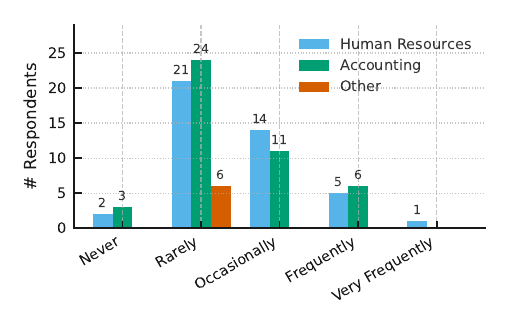}
        \caption{Frequency of encountering cybersecurity threats}
        \label{fig:frequency}
    \end{subfigure}

    \caption{Survey results comparing departments in terms of cybersecurity threat awareness. Note that, n=43 for HR, n=44 for accounting, and n=6 for ``other''.}
    \label{fig:threat-awareness}
    \vspace{-3mm}
\end{figure*}

\subsubsection*{Data Analysis}
We analyzed survey data using a combination of descriptive and inferential statistics, which is consistent with the best practices for quantitative analysis in mixed-methods research \cite{klassen2012best}. This approach enabled us to extend, validate and generalize the findings from the qualitative phase. First, responses from different companies were merged into a single dataset. Entries with less than 90\% completion were excluded to ensure data quality. Matrix and multiple-choice responses were converted to long format, and Likert-scale items were mapped to a 1–5 numerical scale for statistical comparison. Descriptive statistics were computed to identify trends in perceptions, training engagement, and threat awareness. This included means and standard deviations for Likert items and frequency distributions for categorical variables. To assess differences between departments, a series of independent-samples t-tests were conducted to compare responses from HR and accounting participants. T-tests were applied only to Likert-scale items with sufficient coverage, which were deemed appropriate based on group size and the approximate normality of the responses. These tests evaluated whether the observed mean differences between departments were statistically significant, particularly with regard to perceived threats, training preferences, and organizational support.

\subsection{LinkedIn Surveys}
\label{ssec:method3}
\noindent
In addition to company surveys, we also shared our survey on the LinkedIn social network, targeting once again HR and accounting personnel. The LinkedIn response group ($n=10$) served as an external reference point and control group. We used it to assess whether the insights gathered from company-affiliated participants were consistent with the responses of a more diverse and open population. To evaluate potential differences, we performed independent sample t tests to compare responses from the LinkedIn group with those from company-based participants. These comparisons allowed us to examine whether threat perception, training preferences, and attitudes toward cybersecurity awareness varied significantly between organizational and non-organizational contexts. 

\section{Results}
\label{sec:results}

\noindent
We first outline specific threats faced by HR and accounting departments, comparing them to ``other'' departments 
(§\ref{ssec:results_1}). Then, we discuss the main topics of cybersecurity awareness campaigns for these departments (§\ref{ssec:results_2}). Lastly, we detail the key delivery methods (§\ref{ssec:results_3}). These are all reported based on our mixed-methods approach that combines our qualitative semi-structured interviews with our quantitative survey. We will discuss the LinkedIn results in the next section.

\subsection{RQ1: Threats Faced by HR and Accounting}
\label{ssec:results_1}

\noindent RQ1 focuses on elucidating the specific cybersecurity threats perceived by personnel of HR and accounting departments.

\subsubsection*{\textbf{Interviews}}
The transcripts revealed clear threat patterns. Interviewees from HR consistently identified phishing attacks as the most threatening attack vector. HR interviewees specifically stressed the \textit{phishing attempts using job applications} and \textit{requests for employee data as a pretext}. These targeted emails aimed to exploit HR’s administrative processes, especially in recruitment, wherein large volumes of external email communication take place. Importantly, the interviews also revealed that attackers use \textit{job application cover letters and resumes as vectors for malware delivery}, which are usually \quotes{opened} without prior thinking or considering that the file could be malicious.\textit{Impersonation and other types of social engineering attacks} were also commonly reported in the HR context. Attackers frequently attempted to \textit{pose as company executives} by using authority and urgency to pressure HR personnel into disclosing sensitive information or processing fraudulent requests. Such attempts often occur when strict verification procedures are not applied, which leaves HR staff vulnerable to deception.

Accounting interviewees described a different threat landscape. The most frequently mentioned risks included \textit{invoice fraud}, \textit{credential theft} and \textit{unauthorized access to financial systems}. Participants highlighted that attackers often \textit{manipulate invoice templates} or \textit{alter payment details} with the goal of redirecting funds. These attacks are typically executed during high-pressure periods such as monthly or quarterly financial closings, during which employees are under time pressure, impairing thorough verification. Credential-based threats, including \textit{spear-phishing emails} targeting login credentials for accounting or ERP systems, were also described as becoming more sophisticated and dangerous. 


A common topic in both departments was that awareness of general threats exists and it is being addressed in the existing campaigns. However, the main problem is that implementation of effective countermeasures -- especially in day-to-day processes -- remains problematic. This indicates that the personnel are being trained to acknowledge threats, but not how or with what to counter them. Additionally, participants stated that existing awareness campaigns rarely reflect department-specific attack vectors or scenarios. Instead, they always rely on generic examples that do not align with the operational realities and specific threats.

 \begin{figure*}[!t]
    \centering

    \begin{subfigure}[t]{0.32\textwidth}
        \centering
        \includegraphics[width=\linewidth]{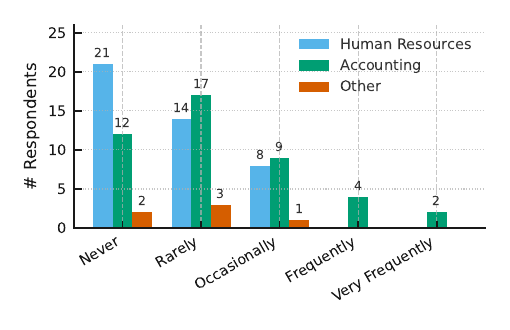}
        \caption{File downloads from unapproved sources}
        \label{fig:downloads}
    \end{subfigure}
    \hfill
    \begin{subfigure}[t]{0.32\textwidth}
        \centering
        \includegraphics[width=\linewidth]{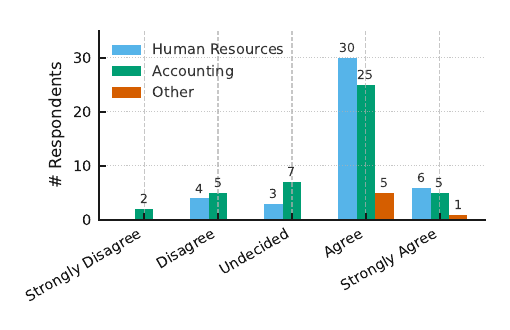}
        \caption{Awareness of third-party risks}
        \label{fig:thirdparty}
    \end{subfigure}
    \hfill
    \begin{subfigure}[t]{0.32\textwidth}
        \centering
        \includegraphics[width=\linewidth]{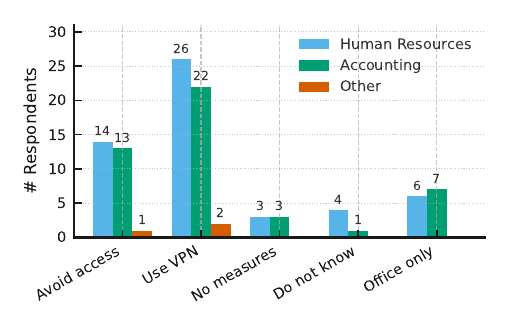}
        \caption{Handling work access over public Wi-Fi}
        \label{fig:wifi}
    \end{subfigure}

    \caption{Comparison of cybersecurity-related practices across departments. Note that, n=43 for HR, n=44 for accounting, and n=6 for ``other''.}
    \label{fig:behavior_comparison}
\end{figure*}

\subsubsection*{\textbf{Surveys}}

In the survey, respondents were asked to select and rank cybersecurity threats most relevant to their work environment. Among HR participants (\texttt{$n=43$}), phishing attacks were stated by 91\% of employees as the primary concern. This was followed by data protection risks (79\%) related to handling personal information. The high percentage confirms that HR employees are aware of their exposure to attacks aimed at data exfiltration or manipulation. However, they are less informed about specific threats such as malicious QR codes, deepfake videos and supply chain attacks. This pattern is also visible in the phishing threat recognition chart (see Figure \ref{fig:phishing}), where HR employees identified phishing threats more often than accounting employees.

For accounting respondents (\texttt{$n=44$}), invoice fraud emerged as the most frequently selected threat (86\%), followed by credential theft (82\%). These results strongly align with the qualitative insights, which identified manipulation of invoices and unauthorized access to financial systems as key concerns in the accounting domain. Moreover, 63\% of accounting respondents also cited ransomware as a relevant threat, particularly in connection with the potential encryption of payment systems or accounting databases. Interestingly, as shown in the chart on confidence in identifying threats (see Figure \ref{fig:confidence}), accounting employees reported lower confidence in recognizing cybersecurity threats compared to HR. This is concerning, given their \quotes{regular} exposure to high-risk attacks.


When asked whether current training materials and awareness measures addressed department-specific risks, only 27\% of the total sample agreed. Among those who disagreed, the majority came from HR and accounting, which confirms that current training lacks contextualization. The participants said the training was too general, too technical, or irrelevant to their real-world challenges. Figure \ref{fig:frequency} shows that HR and accounting employees report facing cyber threats often, as well as their dissatisfaction with the current training. This gap makes it clear that specific departments need awareness campaigns that are more customized to their workflows. Further analysis showed that this dissatisfaction was not limited to a particular level of seniority. Both staff-level employees and managers in HR and accounting indicated that the current cybersecurity training frameworks lack specificity.

\begin{cooltextbox}
    \textsc{\textbf{Answer to RQ1:}}
    HR and Accounting departments face distinct cybersecurity threats tied to their operational workflows. HR is most exposed to phishing and impersonation attacks exploiting hiring processes and organizational hierarchy. Accounting primarily faces invoice fraud, credential theft, and ransomware, especially during financial closings. While both departments recognize general threats, interviews and surveys reveal that current awareness efforts are too generic and fail to address department-specific risks.
\end{cooltextbox}

\begin{figure*}[!t]
    \centering
    \begin{subfigure}[t]{0.32\textwidth}
        \centering
        \includegraphics[width=\textwidth]{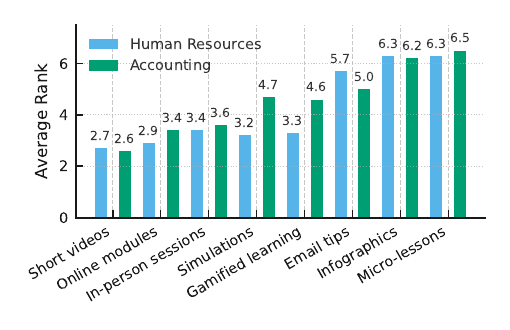}
        \caption{Preferred training format (lower rank is better)}
        \label{fig:training_format}
    \end{subfigure}
        \hfill
    \begin{subfigure}[t]{0.32\textwidth}
        \centering
        \includegraphics[width=\textwidth]{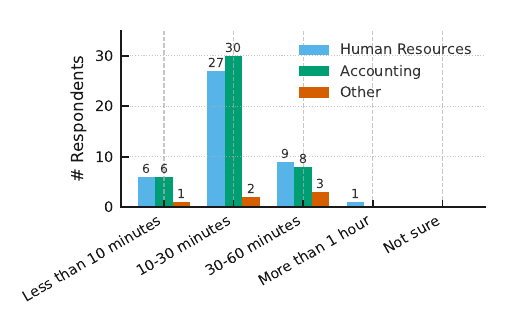}
        \caption{Preferred duration of cybersecurity training}
        \label{fig:training_duration}
    \end{subfigure}
    \hfill
    \begin{subfigure}[t]{0.32\textwidth}
        \centering
        \includegraphics[width=\textwidth]{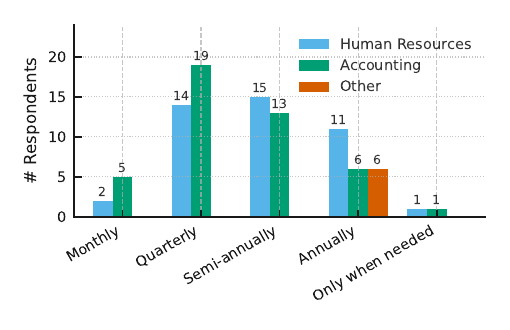}
        \caption{Preferred frequency of cybersecurity training}
        \label{fig:training_frequency}
    \end{subfigure}

    \caption{Training preferences across departments. Note that, n=43 for HR, n=44 for accounting, and n=6 for ``other''.}
    \label{fig:training_preferences}
\end{figure*}

\subsection{RQ2: Main Topics of Awareness Campaings}
\label{ssec:results_2}

\noindent The focus of RQ2 is on identifying the key cybersecurity topics that should be addressed in awareness campaigns to effectively support the specific departmental needs.

\subsubsection*{\textbf{Interviews}}

The interviews revealed that awareness campaigns must be grounded in the operational realities of each role. For HR staff, a central topic should be the secure handling of personal and sensitive employee data. Several HR participants reported uncertainty about which channels and formats are appropriate for transmitting documents such as CVs and contracts. One HR interviewee admitted: \textit{Sometimes I send a CV by email and then I think, maybe I should have encrypted it---but it is not clear}. Another was unsure whether sharing CVs over Microsoft Teams was secure or compliant with company policies.


In accounting, the interviews pointed to a need for awareness content that addresses fraudulent financial requests and manipulation of payment processes. Interviewees described encounters with fake invoices or supplier impersonation. One interviewee stated: ``\textit{It looked exactly like the real invoice -- same layout, same logo -- just a different IBAN}.'' To be effective, awareness campaigns should cover how to verify vendor information, authenticate requests for bank detail changes, and recognize red flags in financial documentation. 

Across both departments, there was uncertainty regarding how to respond to \quotes{suspicious} situations. Several HR and accounting employees noted that they could often identify something \quotes{strange} in a message or attachment, but were unsure what to do. Hence, 
campaigns should clearly communicate reporting procedures and provide clarity on when an issue should be flagged, paused, or passed to security teams. 



\subsubsection*{\textbf{Surveys}}

The survey showed that employees share confidential files such as payroll records or contracts via email or internal messaging tools such as Teams. These methods often lack encryption or audit trails, which makes them vulnerable to man-in-the-middle or misuse attacks. This underscores the need for campaigns to cover practical guidance on secure file transfer, internal data handling policies, and the risks of casual sharing practices. In addition to file sharing, downloading behavior emerged as another weak spot (see Figure~\ref{fig:downloads}). The participants acknowledged retrieving files or templates from unapproved external websites, which can expose systems to malware or data exfiltration. Training materials need to move beyond abstract warnings and instead demonstrate how malicious files are embedded in realistic formats such as fake invoices or HR templates. In accounting, where macro-enabled spreadsheets or automated payment tools are common, awareness campaigns must directly address the threat of malicious documents and promote safe alternatives.

Password management also stood out as a concern. Employees admitted to reusing passwords or storing them in insecure ways. These practices place department-specific systems and tools at risk---especially those that control access to financial data or employee records. Awareness efforts should focus on promoting practical password hygiene with relatable examples. For example, showing how a reused password could compromise both the HR database and payroll systems is considered more useful than merely recommending a theoretical \quotes{strong password} rule. Finally, the survey also showed limited awareness of the cybersecurity risks tied to third-party tools (see Figure~\ref{fig:thirdparty}) and use of public Wi-Fi (see Figure~\ref{fig:wifi}). Campaigns should therefore incorporate content about shadow IT, supplier verification, and the risks of working with unapproved vendors.


\begin{cooltextbox}
    \textsc{\textbf{Answer to RQ2:}}
    Effective cybersecurity awareness campaigns for HR and accounting should focus on secure document handling, invoice fraud scenarios, password hygiene, and clear incident reporting procedures. Interviews emphasized the need for guidance on sharing sensitive files like CVs and contracts, and verifying financial requests. Survey data confirmed risky behaviors, including unsafe downloads, reused passwords, and uncertainty about reporting threats. 
\end{cooltextbox}

\subsection{RQ3: Delivery Methods}
\label{ssec:results_3}

\noindent 
RQ3 focuses on determining which delivery methods are (perceived as being) most effective for engaging employees in cybersecurity awareness efforts.

\subsubsection*{\textbf{Interviews}}

The results obtained from interviews indicate a strong and consistent preference for cybersecurity training methods that are concise, engaging, and adaptable to individual learning needs. Participants highly prefer scenario-based formats over traditional approaches dominated by static text or lengthy slide presentations. One delivery method that stood out above all others was video-based training. It is preferred for its ability to break down complex topics, present realistic and relatable scenarios, and maintain attention through clear pacing and engaging storytelling. Videos can also be revisited as needed, supporting flexible and self-paced learning. As one interviewee explained: \textit{... and I think when you then do a bit of a mixture between a video and slides or you have to click again something or then say what is the right answer. I think that normally works quite well, because it is not... you are not there only passively as well.} 

Accounting department participants in particular highlighted the value of hands-on, decision-driven learning. Formats such as branching scenarios or phishing simulations were favored, as they provided opportunities to apply judgment in situations that closely mirrored real-life threats. These simulations proved to be memorable and relevant, while also often reflecting the different types of fraud attempts and transactional risks. The emphasis was not just on recognizing threats but on practicing how to respond---participants mentioned traditional modules failed to deliver this aspect.

In HR, the need for customizable and time-sensitive learning was a recurring theme. HR professionals noted that their diverse responsibilities require flexibility. They appreciated training that allows them to bypass familiar content so they can spend more time on novel materials. Scenario-based learning was often mentioned, especially when it involved typical HR tasks such as handling confidential employee data or managing payroll communications. 

Cybersecurity staff, meanwhile, were especially critical of conventional training formats such as long webinars or dense policy documents, which they said are often ignored or skimmed by employees. Instead, they advocated for dynamic, visually engaging materials such as short videos, interactive modules, and gamified quizzes. These formats were seen to be more efficient with modern attention spans and reflective of the fast-changing nature of cyber threats.

\subsubsection*{\textbf{Surveys}}

Survey responses confirmed a strong preference for awareness campaigns that are directly relevant to employees' day-to-day tasks. Figure \ref{fig:training_format} presents the average ranking results across departments regarding their preferred methods. Short online videos emerged as the top choice overall, especially among accounting and human resources staff. Participants appreciated these formats for their clarity, visual engagement, and ease of learning. Scenario-based simulations and interactive e-learning modules also consistently ranked in the top three, reflecting a strong inclination toward immersive and participatory formats. In contrast, more static and traditional methods, including email newsletters, infographics, posters, and standalone microlearning sessions, ranked lower. 

For training frequency and duration, employees opted for moderate and time-efficient engagement. When asked how often they would ideally participate in cybersecurity training (see Figure \ref{fig:training_duration}), the most selected were quarterly ($n=36$) and semi-annually ($n=33$). Annual training sessions received moderate support ($n=24$), while monthly training ($n=7$) and ad hoc sessions ($n=2$) were the least favored. A similar pattern emerged regarding training length. Figure \ref{fig:training_frequency} shows that the majority of respondents ($n=64$) preferred sessions lasting between 10 and 30 minutes, with $n=15$ favoring ultra-short formats under 10 minutes. Only a small group supported longer sessions, with $n=16$ preferring 30–60 minutes and just $n=5$ opting for durations exceeding one hour.


The survey also explored whether employees supported the idea of skipping content they already understood. A majority of respondents ($n=65$) expressed strong support for adaptive learning, which shows a widespread desire for training formats that allow learners to focus on unfamiliar or evolving topics. Personalized experiences that respect existing knowledge and streamline the learning process were viewed as more efficient and more motivating.

\begin{cooltextbox}
    \textsc{\textbf{Answer to RQ3:}} Our user studies reveal a clear preference for concise, engaging, and job-relevant cybersecurity training. Employees across departments favored short video-based content, scenario-driven simulations, and interactive modules over static formats like email tips or policy slides. Accounting staff valued practical simulations that reflect phishing and fraud risks, while HR respondents appreciated flexible, self-paced learning tailored to their diverse responsibilities. Overall, participants preferred periodic sessions lasting no more than 30 minutes, and supported adaptive learning options that allow skipping familiar content.
\end{cooltextbox}

\section{Discussion}
\label{sec:discussion}
\noindent
We draw lessons learned from our research (§\ref{ssec:discussion1}), position our work within related studies (§\ref{ssec:discussion2}), and evaluate results of industrial surveys with additional LinkedIn surveys (§\ref{ssec:discussion3}).

\subsection{Key Findings}
\label{ssec:discussion1}

\noindent
We identify four key findings from our results.

The first key finding is that employees in both HR and accounting departments encounter cyber threats that are not only frequent but highly contextual. Our results confirm that phishing remains the dominant threat vector; however, its manifestation differs across domains. In HR, attackers often embed malware in seemingly legitimate job application materials such as CVs and cover letters, knowing these documents are often opened without technical vetting. These attack vectors exploit both human routine and process design. In contrast, accounting departments face a high prevalence of invoice fraud attempts, in which fake payment details are inserted into cloned invoice templates. These threats bypass traditional technical filters because they appear plausible in context, especially during busy periods such as month-end closings. We recommend that future research and campaigns do more to surface and simulate these highly context-driven threats.

The second key finding is the persistent gap between reported security practices and actual behavior. Interviews revealed a strong self-perception of security awareness, wherein employees claimed to follow best practices in areas such as password use and document handling. However, survey data exposed inconsistencies: a considerable number of employees admitted to reusing passwords, sharing files via unsecured platforms, or bypassing encryption when sending sensitive data. These behaviors introduce significant risk, particularly in departments with privileged access to personal or financial records. This divergence suggests the presence of social desirability bias in qualitative self-reporting and highlights the need for behavioral validation mechanisms, such as usage telemetry or targeted simulations. However, another explanation for these inconsistencies is that the overall (perceived) security awareness of the employees that participated in the interviews (all belonging to our case company) was higher w.r.t. the employees that participated in the survey (whose respondents belong to various organizations). 

The third key finding is that traditional training formats are poorly suited to fostering lasting engagement. Our data show that employees overwhelmingly prefer short, video-based content and interactive scenarios tailored to their daily tasks. Delivery formats entailing long sessions and static content, such as slide decks or newsletters, were consistently rated as least effective. The participants cited them as contributing to disengagement and information fatigue. Interestingly, content relevance was ranked as a more powerful motivator than managerial encouragement. This indicates that awareness efforts should prioritize scenario authenticity and personal applicability over hierarchical delivery or generic messaging. Future research should explore adaptive content delivery models that align training with specific job functions and user knowledge baselines.

The fourth key finding centers on the underestimated risk posed by emerging threats and the widespread demand for training personalization. Threat types such as deepfake impersonations, malicious QR codes, and supply chain intrusions were largely unfamiliar to employees, despite their increasing use in targeted attacks. In parallel, over 60\% of respondents indicated a desire to skip content they already understood—suggesting that current one-size-fits-all campaigns not only fail to educate on new threats, but also waste time repeating known material. Thus, these findings call for awareness campaigns that integrate threat intelligence updates with learner autonomy. We advocate for future work on designing modular, updatable training ecosystems that adapt to evolving risks while respecting users’ existing competencies.

\subsection{Comparison to Prior Work}
\label{ssec:discussion2}

\noindent
Previous studies in the domain of cybersecurity awareness, such as those by Alshaikh et al. \cite{alshaikh_applying_2021}, Rege et al. \cite{rege_social_2020}, and He et al. \cite{he_effects_2017}, have emphasized the importance of tailoring security training to organizational contexts and user roles. These works focused on general role-based or sectoral adaptations but did not explore tailoring at the department level. Our findings reveal that the perceived relevance of training content varies sharply between departments within the same organization. Specifically, the results from our interviews and survey show that employees in accounting and HR are exposed to distinct threat types and expect different training formats and content. This observation extends the scope of prior work by demonstrating that department-specific tailoring is necessary to increase engagement and practical relevance.

Nicholson et al. \cite{nicholson_introducing_2018} and Carella et al. \cite{carella_impact_2017} mentioned that training should be user-centered and participatory. However, these proposals are either theoretical or abstract. In contrast, our results identify actionable paths for participatory tailoring by embedding feedback processes at the departmental level. Interviews revealed that the absence of direct communication between campaign designers and individual departments leads to ineffective or misaligned content. Our study contributes new empirical evidence showing that tailoring efforts cannot succeed without a structured and continuous exchange between content creators and targeted department members.

Sterk and Heinemann \cite{sterk_it_2021}, Anjaria and Mishra \cite{anjaria_neuroevolution_2017}, and Schöni et al. \cite{schoni_you_2024} have highlighted the need to align training topics with employees’ existing knowledge and operational responsibilities. However, most existing research does not provide practical guidance for selecting content that reflects department-specific tasks. Our findings address this gap. For example, HR staff expressed a need for training on handling falsified applications and protecting personal data, while accounting employees emphasized phishing, invoice fraud, and data leakage risks. The observed differences in topic preference and perceived preparedness suggest that a standardized curriculum is insufficient.

Alkhazi et al. \cite{alkhazi_assessment_2022}, Reeves et al. \cite{reeves_generic_2023}, and Chowdhury et al. \cite{chowdhury_personalized_2023} have previously noted that generic training formats often fail to engage users, yet few studies have examined the role of perceived relevance as a mediating factor in campaign success. Our findings show that this perception is critical. Participants consistently reported that overly generic content lacked practical value, whereas scenario-based content rooted in their departmental context was seen as useful and memorable. This aligns with the behavioral insights discussed by Khando et al. \cite{khando_enhancing_2021}, who emphasize the importance of personal and contextual alignment in changing security behavior. 

\subsection{Verification Using LinkedIn Surveys}
\label{ssec:discussion3}

\noindent
The LinkedIn survey group that was used as an external reference point provided us with insights regarding the applicability of the study’s findings beyond the participating organizations. Although the sample size was small ($n = 10$), responses were largely consistent with those from participants from companies in terms of threat perception, delivery preferences and training attitudes. The only statistically significant deviation observed through independent sample t-tests concerned the perception of emerging threats. LinkedIn respondents stated that threats were more prominent. This may reflect heightened general awareness or differing exposure across industries. 



\section{Ethical Considerations and Limitations}
\label{sec:ethical}

\noindent We outline the ethical precautions (§\ref{ssec:ethical1}) we followed during our research, and then discuss the limitations of our work (§\ref{ssec:limitations}).

\subsection{Ethical Considerations}
\label{ssec:ethical1}

\noindent 
Our institutions did not have, nor required, a formal IRB process when we carried out our study. Yet, we followed established ethical guidelines throughout our research~\cite{bailey2012menlo}.

\subsubsection*{\textbf{Generic Approach}}

The case company designated a contact person within the cybersecurity department to serve as the primary liaison for all coordination, approvals and communication. This individual helped us ensure that all procedures related to participant recruitment, internal communication and organizational transparency were corresponding with the company’s internal policies and privacy expectations. Ethical safeguards were embedded into every stage of the research process, covering both data collection and dissemination. In addition, the study followed the university’s ethical research standards and adhered to all applicable regulations on data privacy, informed consent, and participant rights. At all stages, particular care was taken to ensure that participants were treated with respect and that their privacy was protected. Moreover, our participants know our identity so they can contact us if they wish their data to be deleted.

The participation in both the qualitative and quantitative part of the study was strictly voluntary. Participants were informed about the academic nature of the research, objectives and the confidentiality of the responses. No pressure or incentives were used to influence participation, we did not use deception, and participation in our study cannot lead to any harm to our participants. For the interview phase, participants received an introductory briefing, both verbally and in writing, at the beginning of each session. They were explicitly informed that their participation was voluntary and that they could withdraw from the interview or decline to answer any question at any point, without any negative consequences. All participants gave us their informed consent prior to data collection.

No data we collected was shared with any third party---not even with the case company: this ensured that our participants were free to express their own thoughts. Indeed, we only provided a high-level summary of the findings focused on thematic insights, which could be used to improve their own cybersecurity posture, but without revealing any individual-level feedback or any raw data. This ensured that employees’ participation could not be traced or evaluated internally.

\subsubsection*{\textbf{Interview Conduction}}

Semi-structured interviews were conducted remotely using MS Teams (\url{https://microsoft.com/en-us/microsoft-teams/}). This solution ensured flexibility for participants while respecting organizational policies and schedules. Audio recording and transcription (to which participants agreed) leveraged MS Teams' built-in functions and they were stored securely on university-managed systems. The transcripts were manually reviewed and anonymized, and all references to names, locations, or other personally identifiable information were removed. Each transcript was assigned a pseudonymous identifier to ensure traceability during analysis without compromising confidentiality.

\subsubsection*{\textbf{Survey Conduction}} 
The survey was carried out using \quotes{Findmind}, which is a Swiss cloud-based survey tool with servers located in Switzerland. The tool operates in compliance with European data protection regulations. Participants accessed the survey through a link distributed through email. At the beginning of the survey, participants were presented with an introductory statement that clarified the purpose of the study, confirmed the anonymity of all responses, and provided contact information for questions or concerns. The survey collected no personally-identifiable or sensitive data~\cite{sensitiveEU,sensitiveUSA}. 

\subsection{Limitations and Threats to Validity}
\label{ssec:limitations}

\noindent 
While this research contributes actionable insights into the tailoring of department-specific cybersecurity awareness campaigns, several limitations should be acknowledged.

The literature search was executed using a defined set of academic databases and a predefined set of keywords. Thus, there is a possibility that some relevant studies may have been omitted. For instance, our primary search terms focused on the concept of \quotes{awareness}, which may have led to the exclusion of studies using adjacent or synonymous terms such as \quotes{training}, \quotes{education}, or \quotes{user engagement}. This poses a potential threat to the completeness of our SLR. To mitigate this, we employed two complementary search strings and performed both backward and forward snowballing. 

Interviews were conducted within a \textit{single multinational case company} operating in the construction sector. Although the organization provided access to critical departments, the findings are embedded within this specific organizational culture, structure and maturity level. Thus, the nature of the findings may not be directly transferable to other organizations, especially those from a different type of industry.

The coding and extraction of papers and interview transcripts were guided by a predefined data extraction sheet, which involved qualitative analysis. Hence, the overall process may be subject to \textit{researcher bias in interpretation of findings}. We addressed this by conducting an independent review by a second researcher. Additionally, the inter-coder reliability score was calculated, which demonstrated a high agreement rate between the two coders, following a similar dual-coding approach as adopted in prior transparency studies~\cite{pekaric2025weprovide}. 

Both phases of the study, including interviews and surveys, relied on voluntary participation. While participant recruitment aimed to ensure representation across key departments, it is possible that those more interested in or aware of cybersecurity topics were more likely to engage. This may have introduced a \textit{self-selection bias}, potentially skewing results towards higher awareness levels or more favorable attitudes toward training. Furthermore, the \textit{limited sample size} in the qualitative phase, while sufficient for thematic saturation, constrains the ability to explore subgroup-specific phenomena (e.g., role seniority, geography or digital literacy).


Finally, \textit{the empirical data collected in our study is only obtained from two departments}. However, we do not claim generalisability of our findings to other departmental units---which is an exercise we leave to future work. 

\section{Conclusions and Recommendations}
\label{sec:conclusions}

\noindent
This study contributes the first department-specific investigation into how cybersecurity awareness campaigns are perceived and internalized by employees in HR and accounting. We show that the prevailing \quotes{one-size-fits-all} approach to training fails to account for the distinct threat landscapes, appropriate campaign structures, and topic relevance required by these critical departments. Both qualitative and quantitative results converge on the demand for short, engaging formats---especially videos and simulations---that are directly relevant to users’ actual tasks. At the same time, we reveal a widespread underestimation of emerging threats such as malicious QR codes, deepfakes, and supply chain attacks. These threats are increasingly leveraged in real-world campaigns but remain largely absent from current training materials.

Based on these findings, we recommend that cybersecurity campaign designers abandon generalized awareness strategies in favor of targeted training pathways aligned with departmental risks and responsibilities. Campaigns should reflect actual decision-making scenarios such as fraudulent payroll requests in HR or altered invoices in accounting. These should offer interactive elements that promote active learning. Adaptive modules that skip known content, clear reporting procedures, and periodic feedback loops can all help increase the effectiveness and credibility of such programs.

Looking forward, our work lays the foundation for future studies examining campaign effectiveness across additional departments, industries, and cultural settings. A promising direction is to measure the long-term behavioral impact of tailored training. For instance, this can be done through follow-up simulations or technical indicators such as secure file-sharing practices. Additionally, future research should explore the role of decentralized models of engagement, such as departmental \quotes{cyber champions}. To support this, we provide our methodological tools and data structures to encourage replication, extension, and comparative analysis by other researchers and practitioners.

\bibliographystyle{IEEEtran}

{\footnotesize
\bibliography{bibliography} 
}

\appendix
\section{Appendix}
\label{app:first}

\subsection{Systematic Literature Review}
\label{sapp:slr}

Tables \ref{tab:search_results} and \ref{tab:included_papers} summarize the search outcomes across different sources and detail the number of papers included at each phase of the selection process as part of the systematic literature review.

\begin{table}[h!]
\caption{Search Results and Inclusion Details}
\label{tab:search_results}
\centering
\begin{tabular}{lccc}
\toprule
\textbf{Source} & \textbf{Search String} & \textbf{Search Results} & \textbf{Included} \\
\midrule
ACM DL & Long  & 99   & 2  \\
IEEE Xplore         & Short & 361  & 8 \\
USENIX       & Short & 1    & 1  \\
NDSS & Short & 0 & 0  \\
ScienceDirect      & Short & 115  & 4  \\
Springer       & Long  & 175  & 1  \\
AIS                 & Long  & 468  & 3  \\
\midrule
\textbf{Total}      &       & 1219 & 21 \\
\bottomrule
\end{tabular}

\end{table}

\begin{table}[h!]
\caption{Number of included papers by search phase}
\label{tab:included_papers}
\centering
\begin{tabular}{lc}
\toprule
\textbf{Phase} & \textbf{Number of included papers} \\
\midrule
Keyword Search    & 21 \\
Backward Search    & 2  \\
Forward Search     & 4  \\
White Literature    & 2  \\
\midrule
\textbf{Total}     & 29 \\
\bottomrule
\end{tabular}

\end{table}

\begin{table*}
\caption{Survey Questions and Answers}
\label{tab:survey_questions_answers}
\centering
\scriptsize
\begin{tblr}{
  width = \linewidth,
  rowsep = 0.01pt,
  colsep = 0.1pt,
  colspec = {Q[m,c,0.4cm]Q[m,c,0.4cm]Q[m,c,3.1cm]Q[m,l,8.3cm]Q[m,l,4.8cm]},
  hlines,
  vlines,
  hline{1,2} = {-}{0.08em},
}
\textbf{ID} & \textbf{RQ} & \textbf{Category} & \textbf{Question English} & \textbf{Answers} \\
1 & - & Demographics and Context & What department do you currently work in? & HR, Finance, Other \\
2 & - & Demographics and Context & What is your role within your department? & Employee, Team Lead, Manager, Director (or above) \\
3 & - & Demographics and Context & How long have you worked in your current role? & Less than 1 year, 1-3 years, 4-7 years, 8-10 years, More than 10 years \\
4 & - & Demographics and Context & How old are you (age category)? & 18-24, 25-34, 35-44, 45-54, 55-65 \\
5 & 3 & Current Training & How often do you currently receive cybersecurity awareness training (e.g. online courses, workshops)? & Monthly, Quarterly, Twice per year, Once per year, I don't recall receiving any \\
6 & 3 & Current Training & How long does your typical cybersecurity training session last currently? & Less than 10 minutes, 10-30 minutes, 30-60 minutes, More than 1 hour, Not sure \\
7 & 3 & Current Training & The current cybersecurity training format matches how I learn best. & Likert \\
8 & 2 & Current Training & I have changed how I handle emails/files due to past cybersecurity training. & Likert \\
9 & 1 & Knowledge and Threat Awareness & I believe my department is more targeted than others when it comes to cybersecurity threats. & Likert \\
10 & 2 & Knowledge and Threat Awareness & How relevant do you currently find cybersecurity awareness materials provided by your organization (to my department’s daily tasks)? & Likert \\
11 & 1 & Knowledge and Threat Awareness & The cybersecurity training I receive helps me feel confident in spotting and responding to threats. & Likert \\
12 & 2 & Knowledge and Threat Awareness & How confident are you in identifying cybersecurity threats that are specific to your department (e.g., payroll scams, invoice fraud)? & Likert \\
13 & 2 & Knowledge and Threat Awareness & I can recognize threats such as phishing emails targeting my departments activities (e.g., fake payroll or invoice requests). & Likert \\
14 & 1 & Knowledge and Threat Awareness & How often do you encounter cybersecurity threats (e.g., phishing attempts)? & Never, Rarely, Occasionally, Frequently, Very Frequently \\
15 & 2 & Knowledge and Threat Awareness & I am aware of the risks associated with third-party platforms/tools we use. & Likert \\
16 & 2 & Knowledge and Threat Awareness & How often do you download files or templates from websites not provided by IT or your team? & Never, Rarely, Occasionally, Frequently, Very Frequently \\
17 & 2 & Knowledge and Threat Awareness & When setting a password for a work-related account, which of the following do you typically consider? & Length, Symbols, Reuse, Remembering ease, System-generated \\
18 & 2 & Knowledge and Threat Awareness & Which of these statements best describes how you manage your work passwords? & Reuse same, Vary slightly, Different for each, Use password manager \\
19 & 2 & Knowledge and Threat Awareness & When using a public Wi-Fi (e.g., airport or café) how do you handle work-related access? & Avoid, Use VPN, Use without measures, Don't know, Never work outside \\
20 & 2 & Knowledge and Threat Awareness & I know what steps to take after a suspected cyber security breach. & Likert \\
21 & 2 & Knowledge and Threat Awareness & Do you know how to report a cyber security incident within your organization? & Likert \\
22 & 2 & Knowledge and Threat Awareness & I know the correct steps to securely store sensitive work-related documents/data. & Likert \\
23 & 2 & Knowledge and Threat Awareness & How do you usually share confidential files (e.g., payroll, contract data) with colleagues or vendors? & Secure internal tools, Regular email, Cloud services, Not sure \\
24 & 2 & Knowledge and Threat Awareness & Cyber security training at my company addresses newly emerging threats in a timely manner. & Likert \\
25a & 2 & Knowledge and Threat Awareness & I am familiar with ... deepfake videos as a cybersecurity threat. & Yes/No \\
25b & 2 & Knowledge and Threat Awareness & I am familiar with ... malicious QR codes as a cybersecurity threat. & Yes/No \\
25c & 2 & Knowledge and Threat Awareness & I am familiar with ... supply chain cyber attacks. & Yes/No \\
25d & 2 & Knowledge and Threat Awareness & I am familiar with ... phishing via messaging apps or social media. & Yes/No \\
26 & 3 & Delivery Method Preferences & I feel comfortable asking questions or making mistakes during cybersecurity training. & Likert \\
27 & 3 & Delivery Method Preferences & I prefer cybersecurity awareness campaigns that use realistic scenarios based on my daily tasks. & Likert \\
28 & 3 & Delivery Method Preferences & What would motivate you most to actively participate in cybersecurity training? & Relevance to tasks, Engaging content, Incentives, Manager encouragement \\
29 & 3 & Delivery Method Preferences & Rank the following training formats in order of your preference (1 = most preferred): & Videos, Interactive modules, In-person sessions, Simulations, Games, Newsletters, Infographics, Micro-learning \\
30 & 3 & Delivery Method Preferences & How frequently would you prefer cybersecurity training to occur to maintain awareness effectively? & Monthly, Quarterly, Semi-annually, Annually, Only when needed \\
31 & 3 & Delivery Method Preferences & Based on your selected training frequency, how long should one cybersecurity training session last? & Less than 10 min, 10–30 min, 30–60 min, More than 1 hour, Not sure \\
32 & 3 & Delivery Method Preferences & I would like the ability to skip parts of the training I already know. & Likert \\
33 & 3 & Organizational Culture and Management Support & My team head/manager/director reinforces cybersecurity best practices in our team. & Likert \\
34 & 1 & Organizational Culture and Management Support & In your opinion, how seriously does your organization take cybersecurity? & Likert \\
35 & 1 & Organizational Culture and Management Support & Cybersecurity is discussed regularly within my department (e.g., in team meetings or chats). & Likert \\
36 & 1 & Organizational Culture and Management Support & I feel cybersecurity is mainly the responsibility of the IT department and not part of my daily work experience. & Likert \\
\end{tblr}

\end{table*}

\subsection{Interview and Survey Guides}
\label{sapp:data_collection_instruments}
\noindent
Table \ref{tab:survey_questions_answers} presents the questions as well as possible answers from the company survey. In addition, in Table \ref{tab:interview_guide_departments}, we provide the interview guide with all the questions targeting cybersecurity specialists.

\begin{table}[htbp]
\caption{Interview guide for cybersecurity specialists}
\label{tab:interview_guide_cyber_specialists}
\footnotesize
\centering
\noindent \footnotesize
\begin{tabularx}{\columnwidth}{|c|c|c|X|}
\hline
\textbf{ID} & \textbf{RQ} & \textbf{Category} & \textbf{Question} \\
\hline
1 & - & Warm-Up & What is your current role? \\
\hline
2 & - & Warm-Up & How many years of experience do you have in your role? \\
\hline
3 & - & Warm-Up & How old are you? \\
\hline
4 & - & Warm-Up & What topics are you working on the most? \\
\hline
5 & 1 & General & Based on your expertise, what are the most common cybersecurity threats faced by organizations? \\
\hline
6 & 1 & General & What role do human errors play in cybersecurity risks? \\
\hline
7 & 1 & HR & What are the top three cybersecurity risks associated with HR departments? \\
\hline
8 & 1 & HR & How do cyber criminals typically exploit vulnerabilities in HR-departments? \\
\hline
9 & 1 & Finance & What are the top three cybersecurity risks associated with Finance departments? \\
\hline
10 & 1 & Finance & How do cyber criminals typically exploit vulnerabilities in Finance departments? \\
\hline
11 & 2 & Recommendations & What preventative measures do you recommend for addressing department-specific cybersecurity threats? \\
\hline
12 & 2 & Recommendations & What topics should be covered in the awareness campaigns for the departments? \\
\hline
13 & 3 & Recommendations & What delivery method would you think is engaging for the departments? \\
\hline
14 & - & Cool-Down & Is there anything else you would like to add that was not covered in this interview that you think is important? \\
\hline
15 & - & Cool-Down & Do you have a question for me? \\
\hline
\end{tabularx}

\end{table}

\subsection{Survey Demographics}
\label{sapp:demographics}
\noindent In the following, we provide an overview of the demographics of the participants from the company survey. 
Table \ref{tab:role_distribution} showcases the role distribution; Table \ref{tab:tenure_distribution} highlights the tenure of department affiliation; and Table \ref{tab:age_distribution} visualizes the age distribution of the participants.


\noindent Table \ref{tab:role_distribution} categorizes participants by their job roles and department. The majority of respondents in both the HR and Accounting departments were employees (30 and 34, respectively), followed by team leads, managers, and directors. Accounting had slightly higher representation in senior roles, including directors ($n = 3$), than HR ($n = 1$). All participants in the "Other" category identified as employees.
\begin{table}[htbp]
\caption{Role Distribution by Department}
\label{tab:role_distribution}
\centering
\footnotesize
\begin{tabular}{|c|c|c|c|}
\hline
\textbf{Role} & \textbf{HR} & \textbf{Accounting} & \textbf{Other} \\
\hline
Employee & 30 & 34 & 6 \\
\hline
Team Lead & 8 & 4 & 0 \\
\hline
Manager & 4 & 3 & 0 \\
\hline
Director (or above) & 1 & 3 & 0 \\
\hline
\end{tabular}

\end{table}

\noindent Table \ref{tab:tenure_distribution} details participants’ length of service at their current role, categorized by department. Most respondents had between one and seven years of experience. Notably, the accounting department had a higher number of participants in the \quotes{more than ten years} category ($n = 12$). HR participants were more evenly distributed across the lower tenure brackets. Minimal responses were recorded for the "Other" category.

\noindent Table \ref{tab:age_distribution} shows the age ranges of participants by department. The largest age group overall was 35–44, followed by 45–54. HR had a broader age distribution, while accounting was more concentrated in the 45–54 age range. Participants aged 18–24 were minimally represented in all groups.

\FloatBarrier
\begin{table}[htbp]
\caption{Tenure Distribution by Department}
\label{tab:tenure_distribution}
\centering
\footnotesize
\begin{tabular}{|c|c|c|c|}
\hline
\textbf{Tenure} & \textbf{HR} & \textbf{Accounting} & \textbf{Other} \\
\hline
Less than 1 year & 5 & 9 & 0 \\
\hline
1-3 years & 17 & 14 & 0 \\
\hline
4-7 years & 12 & 6 & 2 \\
\hline
8-10 years & 6 & 3 & 1 \\
\hline
More than 10 years & 3 & 12 & 3 \\
\hline
\end{tabular}

\end{table}

\FloatBarrier
\begin{table}[htbp]
\caption{Age Distribution by Department}
\label{tab:age_distribution}
\centering
\footnotesize
\begin{tabular}{|c|c|c|c|}
\hline
\textbf{Age} & \textbf{HR} & \textbf{Accounting} & \textbf{Other} \\
\hline
18–24 years & 1 & 3 & 0 \\
\hline
25–34 years & 9 & 7 & 0 \\
\hline
35–44 years & 17 & 14 & 3 \\
\hline
45–54 years & 13 & 18 & 1 \\
\hline
55–65 years & 3 & 2 & 2 \\
\hline
\end{tabular}

\end{table}

\end{document}